\documentstyle[12pt]{article}
  
\begin{document} 
\title{THE ZETA FUNCTION METHOD AND THE HARMONIC OSCILLATOR PROPAGATOR}
\author{{\Large F.A. Barone $^{\star}$, C. Farina$^{\dagger}$} \\\\
Instituto de F\'{\i}sica - UFRJ - CP 68528
\\
Rio de Janeiro, RJ, Brasil - 21945-970.}
\maketitle
\begin{abstract}
{We show how the pre-exponential factor of the Feynman propagator for the harmonic oscillator can be computed by the generalized $\zeta$-function method. Besides, we establish a direct equivalence between this method and  Schwinger's propertime method.}
\end{abstract}

\bigskip
\vfill
\noindent $^{\star }$ {e-mail: fabricio@if.ufrj.br}

\noindent $^\dagger$ {e-mail: farina@if.ufrj.br}

\pagebreak

\noindent 
In a recent paper that appeared in this journal \cite{Holstein} the harmonic oscillator propagator was evaluated by a variety of ways, all of them based on path integrals. In fact, some of them did not involve any explicit computation of the Feynman path integral, but their common starting point was actually an expression for the harmonic oscillator propagator which was explicitly derived  by path integral means, namely (we are using as much as possible the notation of reference \cite{Holstein}):
\begin{equation}
D_F(z_f,t_f;z_i,t_i)=\left({\det {\cal O}\over \det{\cal O}^{(o)}}\right)^{-1/2}
\;\sqrt{{m\over 2\pi i\hbar (t_f-t_i)}}\;\, 
\exp\left\{{i\over\hbar}\, S[z_{cl}]\right\}\; ,
\end{equation}
where 
\begin{equation}
{\cal O}=\omega^2+{d^2\over dt^2}\;\;\;\;\; , \;\;\;\;\;
{\cal O}^{(o)}={d^2\over dt^2}\; ,
\end{equation}
and the determinants must be computed with Dirichlet boundary conditions. In the previous equation $S[z_{cl}]$ means the classical action, that is, the functional action evaluated at the classical solution satisfying the Feynman conditions $z(t_i)=z_i$ and $z(t_f)=z_f$ and the factor before the exponential is the so called pre-exponential factor, which we shall denote by $F(t_f-t_i)$. In reference \cite{Holstein} three distinct methods were presented for the computation of $F(t_f-t_i)$: ({\it i}) it was computed directly by the products of the corresponding eigenvalues of ${\cal O}$ and ${\cal O}^{(o)}$ (some care must be taken here since both products are infinite, but their ratio is finite); ({\it ii}) it was computed with the aid of Schwinger's propertime method \cite{Schwinger51} (an introductory presentation of this method with simple applications can be found in reference \cite{Farina98}); ({\it iii}) it was computed by the Green function approach (a variety of simple examples worked out with this approach can be found in references \cite{Farina94,Farina95}). 

In this note we just add to the previous list one more method for computing the pre-exponential factor of the harmonic oscillator propagator, namely, the generalized $\zeta$ function method, so that this note can be considered as a small complement of Holstein's paper \cite{Holstein}. In fact, every time we make a semiclassical approximation, no matter it is in the context of quantum mechanics or quantum field theory, we will get involved with the computation of a determinant of a differential operator with some boundary conditions. If we try naively to compute these determinants as the products of the corresponding eigenvalues we will get ill defined expressions. Hence, it is imperative to give a finite prescription for computing determinants for these cases. The generalized $\zeta$-function method is precisely one possible way of doing that. It was introduced in physics in the middle seventies \cite{Zeta} and it is in fact a very powerful regularization prescription which has applications in several branches of physics (a detailed discussion can be found in reference \cite{ElizaldeB}). This method, as we will see, is based on an analytical extension in the complex plane. We think that the harmonic oscillator propagator is the perfect scenario for introducing such an important method, because undergraduate students are all familiarized with the quantum harmonic oscillators and besides, it is the first non-trivial example after the free particle. In what follows, we shall first introduce briefly the $\zeta$-function method, then we shall apply it to compute $F(t_f-t_i)$ for the harmonic oscillator propagator and finally, we shall establish a direct equivalence between this method and Schwinger's propertime method.

Consider an operator $A$ and let us assume, without loss of generality, that it has a discrete set of non-degenerate eigenvalues $\{\lambda_n\}$. When there is only a finite number of eigenvalues $\det A$ is just given by the product of these eigenvalues and we can write:
\eject
\begin{eqnarray}\label{definicao1}
\det A&=&\prod_n\lambda_n\nonumber\\
&=& \prod_n\, \exp\left\{{\log\lambda_n}\right\}\nonumber\\
&=& \exp\left\{\sum_n\log\lambda_n\right\}\nonumber\\
&=&\exp\left\{-\sum_n\left({\partial\lambda_n^{-s}\over \partial s}\right)
_{s=0}\right\}\nonumber\\
&=&\exp\left\{-{\partial\zeta\over\partial s}(0;A)\right\}\; ,
\end{eqnarray}
where we defined the generalized zeta function associated to the operator $A$ as
\begin{equation}\label{definicao2}
\zeta(s;A)=Tr\, A^{-s}\; .
\end{equation}

However, when there is an infinite number of eigenvalues (and these are the cases of interest in physics), as it occurs when $A$ is a differential operator, the product of the eigenvalues will be an ill defined quantity and will not serve anymore as a good prescription for 
$\det A$. In other words, expression (\ref{definicao1}) with $\zeta(s;A)$ given by (\ref{definicao2}), as it stands, is meaningless because it is not valid anymore to use that:
\begin{equation}
\sum_n\left({\partial\lambda_n^{-s}\over \partial s}\right)
_{s=0}=\left\{{\partial\over \partial s}\left(\sum_n \lambda_n^{-s}\right)
\right\}_{s=0}\; .
\end{equation}
Hence, we need for these cases to define a finite prescription for $det A$. The generalized zeta function prescription consists basically of the following three steps: {\bf (i)} we first compute the eigenvalues of $A$ subjected to the appropriate boundary conditions and then write down the corresponding $\zeta$-function $\zeta(s;A)=Tr\, A^{-s}=\sum_n\lambda_n^{-s}$ ; {\bf (ii)} Since the last sum does not converge at $s=0$, we make an analytical extension of this function to the whole complex plane of $s$ (or at least to a domain that contains the origin); {\bf (iii)} after the analytical extension is made, we just write $\det A=\exp\left\{-\zeta^\prime(s=0;A)\right\}$.

In order to apply the $\zeta$-function method described above in the computation of $F(t_f-t_i)$ we first need to find the eigenvalues of ${\cal O}$ with Dirichlet boundary conditions.  For convenience, we shall make the rotation in the complex plane  $t=e^{-i\pi/2}\,T=-i\,T$. Let us also define the corresponding finite interval in $T$ by $t_f-t_i=-i(T_f-T_i)=-i\,\tau$. We then have that
\begin{equation}
{\cal O}=\omega^2+{d^2\over dt^2}\;\;\;\longrightarrow\;\;\; 
{\cal O}_T=\omega^2-{d^2\over dT^2}\; .
\end{equation}
This analytical extension guarantees that all the eigenvalues (now of the operator ${\cal O}_T$) are positive. Of course, after the calculations are finished we must undo this  transformation, that is, we must substitute $\tau=i(t_f-t_i)$. Solving the eigenvalue equation ${\cal O}_T\, f_n(T)=\lambda_n\, f_n(T)$ with Dirichlet boundary conditions $f_n(0)=0=f_n(\tau)$ we get:
\begin{equation}
f_n(T)=\left\{\sin\left({n\pi\over\tau}T\right)\, ;\; n=1,2,...
\right\}\;\;\;\; ;\;\;\;\; 
\lambda_n=\omega^2+{n^2 \pi^2\over \tau^2}\; ;\; n=1,2,...
\end{equation}
Consequently, the associated generalized $\zeta$-function is given by
\begin{equation}\label{zeta1}
\zeta(s;{\cal O}_T)=\sum_{n=1}^\infty {1\over \lambda_n^s}
=\left({\tau\over\pi}\right)^{2s}\sum_{n=1}^\infty 
{1\over (n^2+\nu^2)^s}\; ,
\end{equation}
where we defined $\nu=\omega\tau/\pi$. Since the above series converges only for ${\cal R}e\, s>1/2$, we need to make an analytical extension in the complex plane of $s$ to include the origin. However, this can be done with no effort at all, for this series is precisely the so called non-homogeneous Epstein function, which we shall denote simply by $E_1(s;\nu^2)$ and whose analytical extension to the whole complex plane is well known  and is given by \cite{Ambjorn,ElizaldeB} (see the appendix for a brief deduction):

\begin{equation}\label{E1}
E^{\nu^2}(s;1)=-{1\over 2\nu^{2s}}+
{\sqrt{\pi}\over 2\nu^{2s-1}}{\Gamma(s-1/2)\over \Gamma(s)}
+{2\sqrt{\pi}\over \Gamma(s)}\;\sum_{n=1}^\infty\left({n\pi\over \nu}
\right)^{s-1/2} K_{s-1/2}(2n\pi\nu)\; ,
\end{equation}
where $K_\mu(z)$ is the modified Bessel function of second kind. This is a meromorphic function in the whole complex plane with simple poles at 
$s=1/2,\;-1/2,\;-3/2,\;...,$ so that we can take its derivative at $s=0$ without any problem. Substituting the sum appearing on the r.h.s. of (\ref{zeta1}) by the analytical extension given by (\ref{E1}), we may cast $\zeta(s;{\cal O}_T)$ into the form:
\begin{equation}
\zeta(s;{\cal O}_T)=-{1\over 2}\left({\tau\over \pi\nu}\right)^{2s}+{F(s)\over \Gamma(s)}\; ,
\end{equation}
where 
$$
F(s)=\left({\tau\over\pi}\right)^{2s}\left\{ {\sqrt{\pi}\over 2\nu^{2s-1}}\Gamma(s-1/2)+
2\sqrt{\pi}\sum_{n=1}^\infty\left({n\pi\over \nu}\right)^{s-1/2}
K_{s-1/2}(2n\pi\nu)\right\}
$$ 
is analytic at $s=0$. Taking, then, the derivative with respect to $s$ at $s=0$ and using that $\Gamma(s)\approx 1/s$ for $s\rightarrow 0$, we get
\begin{eqnarray}
\zeta^\prime(s=0;{\cal O}_T)&=&-\log\left({\tau/\pi\nu}\right)
+\lim_{s\rightarrow 0}
\left\{ -{\Gamma^\prime(s)\over \Gamma^2(s)}
F(s)+{F^\prime(s)\over \Gamma(s)}\right\}\nonumber\\
&=&
-\log\left({\tau/\pi\nu}\right)+F(0)\; .
\end{eqnarray}
From the above expression for $F(s)$ we readly compute $F(0)$, so that
\begin{equation}\label{zetaprime1}
\zeta^\prime(s=0;{\cal O}_T)=
-\log\left({\tau/\pi\nu}\right)
+\left[ {\sqrt{\pi}\nu\Gamma(-1/2)\over 2}+
2\sqrt{\pi}\sum_{n=1}^\infty \sqrt{{\nu\over n\pi}}
K_{-1/2}(2n\pi\nu)\right]\; .
\end{equation}
Using that $\Gamma(-1/2)=-2\sqrt{\pi}$ and $K_{-1/2}(z)=
\sqrt{\pi/2z}\, e^{-z}$, we obtain
\begin{equation}\label{zetaprime2}
\zeta^\prime(s=0;{\cal O}_T)=\log\left(\pi\nu/\tau\right)
-\pi\nu+\sum_{n=1}^\infty {1\over n}\; e^{-n 2\pi\nu}\; .
\end{equation}
It is not a difficult task to show that the above sum is given by (take its derivative with respect to $\nu$, sum the resultant geometric series and then integrate  in $\nu$; in order to eliminate the arbitrary integration constant, just use the fact that this sum must vanish for $\nu\rightarrow\infty$ ):
\begin{equation}\label{soma1}
\sum_{n=1}^\infty {1\over n}\; e^{-n 2\pi\nu}=
\pi\nu-\log\left[ 2\;\mbox{sinh}(\pi\nu)\right]\; .
\end{equation}
From equations (\ref{zetaprime2}) and (\ref{soma1}) we then have
\begin{equation}\label{zetaprime3}
\zeta^\prime(s=0;{\cal O}_T)=
\log\left[{\omega\over 2\sinh(\omega\tau)}\right]\; ,
\end{equation}
where we used that $\nu=\omega\tau/\pi$. For the operator ${\cal O}_T^{(o)}$ we immediatly get (it suffices to make $\omega\rightarrow 0$ in the previous formula):
\begin{equation}\label{zetaprime4}
\zeta^\prime(s=0;{\cal O}^{(o)}_T)=
\log\left[{1\over 2\tau}\right]\; .
\end{equation}
Collecting all the previous results and rotating back to the real time ($\tau=i(t_f-t_i)$), we finally obtain
\begin{eqnarray}\label{fator}
F(t_f-t_i)&=&\sqrt{{\exp\left[ -\zeta^\prime(0,{\cal O})\right]
\over
\exp\left[ -\zeta^\prime(0,{\cal O}^{(o)})\right]}}
\times\sqrt{{m\over 2\pi i\hbar(t_f-t_i)}}\nonumber\\
\nonumber\\
&=& \sqrt{{m\omega\over 2\pi i\hbar \sin[\omega(t_f-t_i)]}}\; ,
\end{eqnarray}
where we used that $\sinh(i\theta)=-i\,\sin\,\theta$, in perfect agreement with \cite{Holstein}.

Before we finish this note, we think it is interesting to establish a general equivalence between the $\zeta$-function method and the Schwinger's propertime method. From the $\zeta$-function method just presented, we can write
\begin{equation}\label{equiv1}
\log\det\,{\cal O}=-\zeta^\prime(s=0;{\cal O})\; .
\end{equation}
On the other hand, with the aid of the Mellin transform \cite{Arfken} applied here to an operator ${\cal O}$ with positive eigenvalues we can write
\begin{eqnarray}\label{equiv2}
\zeta(s;{\cal O})&=& Tr\, {\cal O}^{-s}\nonumber\\
&=& Tr\, {1\over\Gamma(s)}\int_0^\infty\, d\tau\, \tau^{s-1}\, e^{-{\cal O}\tau}\; .
\end{eqnarray}
However, last expression is not analytic at $s=0$ (though the presence of the exponential guarantees a well behaviour for large $s$, the limit $s\rightarrow 0$ is a divergent one), so that as it stands it is not valid to take the $s$ derivative at $s=0$. In order to circumvent this problem, we make the modification (regularization)
\begin{equation}\label{reg}
\zeta(s;{\cal O})\longrightarrow \zeta(s,\alpha;{\cal O})=
{1\over \Gamma(s)}\int_0^\infty\, d\tau\, \tau^{s+\alpha-1}\, e^{-{\cal O}\tau}\; ,
\end{equation} 
where $\alpha$ is big enough to ensure that the previous expression is well behaved at $s=0$. Hence, taking first the $s$ derivative at $s=0$ and then taking the limit $\alpha\rightarrow 0$ we obtain:
\begin{eqnarray}
\label{equivfinal}
-\zeta^\prime(s=0;{\cal O})&=&-\lim_{\alpha\rightarrow 0}
\lim_{s\rightarrow 0}{\partial\zeta\over\partial s}(s,\alpha;{\cal O})
\nonumber\\
&=&-\lim_{\alpha\rightarrow 0}\lim_{s\rightarrow 0}
Tr\Biggl\{ -{\Gamma^\prime(s)\over\Gamma^2(s)}
\int_0^\infty\, d\tau\, \tau^{s+\alpha-1}\, e^{-{\cal O}\tau}\nonumber\\
&\ &\ \ \ \ \ \ \ \ \ \ \ \ \ \ \ +{1\over \Gamma(s)}\int_0^\infty\; 
d\tau\; \log \tau\, \tau^{s+\alpha-1}\; e^{-{\cal O}\tau}
\Biggr\}\nonumber\\
&=&-\lim_{\alpha\rightarrow 0} Tr
\Biggl\{\int_0^\infty\, {d\tau\over \tau}\, \tau^\alpha\, 
e^{-{\cal O}\tau}\Biggr\}\; .
\end{eqnarray}
Last expression corresponds precisely to Schwinger's formula written in a regularized way. Here we regularized by introducing positive powers of $\tau$, but other regularization schemes can also be used, as for example the one used by Schwinger \cite{Schwinger92} in the computation of the Casimir effect \cite{Casimir48} (for a simple introduction to this effect with some historical remarks see reference \cite{ElizaldeAJP}). It is common to write the above expression formally with  $\alpha=0$, but in fact, before taking this limit one should get rid off all spurious terms (those with no physical meaning). 

In this note we have presented the generalized $\zeta$-function method for computing determinants in a very introductory level. A detailed discussion with a great variety of examples can be found in reference \cite{ElizaldeB}. One of the greatest advantadges of this method is that for almost all differential operators and boundary conditions that are relevant in physics, the corresponding generalized $\zeta$-function (after the analytical extension is made) is a meromorphic function in the whole complex plane which is analytic at the origin. Furthermore, this method can also be applied successfully in many other branches of physics, as for example, statistical mechanics and quantum field theory among others. Of course there are many easier ways of obtaining $F(t_f-t_i)$ for the harmonic oscillator, but our purpose here was to introduce a new method, which is a powerful one and widely used nowadays. In this sense, we think that the harmonic oscillator provided a perfect scenario for the understanding of the three basic steps of the method, since every undergraduate student is somehow familiarized with the harmonic oscillator. 
%
%

\section*{\bf Appendix}

	In this appendix we shall obtain the  analytical extension of the Epstein function $E_{1}^{\nu^{2}}(s;1)$,  given in the text by equation (\ref{E1}). With this goal, we first write down an equation involving the Gamma function, which follows directly from its  definition, namely \cite{Arfken}:
\begin {equation}
\label {defGamma}
\Gamma (z) A^{-z}=\int_{0}^{\infty}d\tau \;\tau^{z-1}\; e^{-A\tau}\ \ ,\ \ 
Re(z)>0
\end {equation}
Using the previous equation with $A=n^2+\nu^2$, the Epstein function can be written in the form:
\begin{eqnarray}
\label {defEpstein}
E_{1}^{\nu^{2}}(s;1)&=&\sum_{n=1}^{\infty}{1\over (n^{2}+\nu^{2})^{s}}
\nonumber\\
&=&{1\over\Gamma(s)}\int_{0}^{\infty} d\tau \;\tau^{s-1}\; 
e^{-\nu^{2}\tau}\sum_{n=1}^{\infty}e^{-n^{2}\tau}\; .
\end {eqnarray}

On the other hand, from the so called  Poisson summation rule \cite{Arfken}, we can write that:
\begin {equation}
\sum_{n=1}^{\infty}e^{-n^{2}\tau}=-{1\over 2}+{1\over 2}{\sqrt{\pi\over \tau}}+
{\sqrt{\pi\over \tau}}\sum_{n=1}^{\infty}e^{-n^{2}\pi^{2}(1/\tau)}
\end {equation}
%
%

Substituting the last into (\ref{defEpstein}), we get:
\begin {eqnarray}
\label {Epsteinintermediario2}
E_{1}^{\nu^{2}}(s;1)&=&{-1\over 2\Gamma(s)}\int_{0}^{\infty}d\tau 
\;\tau^{s-1}e^{-\nu^{2}\tau}+{\sqrt{\pi}\over 2\Gamma(s)}\int_{0}^{\infty}d\tau 
\;\tau^{s-3/2}\;e^{-\nu^{2}\tau}
\nonumber\\
&+&{\sqrt{\pi}\over\Gamma(s)}\sum_{n=1}^{\infty}\int_{0}^{\infty}d\tau 
\;\tau^{s-3/2}\; e^{-\nu^{2}\tau-n^{2}\pi^{2}/\tau}
\end {eqnarray}

Using (\ref{defGamma}), the first and second integrals of the right hand side of the above equation can be written directly in term of Euler Gamma functions. For the last term, we use the integral representation for the modified Bessel function of second kind:

\begin {equation}
\int_{0}^{\infty}dx x^{\alpha-1}x^{-\beta/x-\gamma x}=2\biggl({\beta\over\gamma}\biggr) ^{\alpha/2}
\; K_{\alpha}(2\sqrt{\beta\gamma})\ \ ,\ \ \;\;\; Re\beta , Re\gamma >0\; .
\end {equation}
Therefore, we finally obtain for equation (\ref{Epsteinintermediario2}) that:

\begin {equation}
E_{1}^{\nu^{2}}(s;1)=-{1\over 2\nu^{2s}}+{\sqrt{\pi}\Gamma(s-1/2)\over 
2\Gamma(s)\nu^{2s-1}}+{\sqrt{\pi}\over\Gamma(s)}\sum_{n=1}^{\infty}\biggl( {n\pi\over\nu}\biggr) ^{s-{1\over2}}K_{s-{1\over2}}(2\pi n\nu)\; ,
\end {equation}
\vskip 0.2 cm
\noindent
which is precisely equation (\ref{E1}). Some comments are in order here: {\bf (i)} to say that the previous equation corresponds to the analytical extension of $E^{\nu^2}(s;1)$ to a meromorphic function in the whole complex plane means that this expression is an analytical function in the whole complex plane except by an enumerable number of poles (which can be infinite) and coincides with the original sum in the region where the sum was defined; {\bf (ii)} it is worth emphasizing that the above expression is analytic at the origin; in fact, the structure of simple poles of this function is dictated by the poles of the Euler gamma function. It is easy to see that the poles are located at $s=1/2,$ $-1/2,$ $-3/2,$ $-5/2.$ etc..


\end{document}